\documentclass[superscriptaddress,secnumarabic,twocolumn,amssymb,amsmath,nobibnotes,aps,prd,showkeys,nofootinbib]{revtex4}
\usepackage{}%
\usepackage{graphicx}
\usepackage{epsf}
\usepackage{bm}
\usepackage{amsmath}
\usepackage{amsfonts}
\usepackage{amssymb}
\usepackage{epstopdf}%
\usepackage{color}

\begin{document}

\title{Dynamical Dark sectors and Neutrino masses and abundances}

\author{Weiqiang Yang}
\email{d11102004@163.com}
\affiliation{Department of Physics, Liaoning Normal University, Dalian, 116029, P. R. China}

\author{Eleonora Di Valentino}
\email{eleonora.divalentino@manchester.ac.uk}
\affiliation{Jodrell Bank Center for Astrophysics, School of Physics and Astronomy, University of Manchester, Oxford Road, Manchester, M13 9PL, UK}

\author{Olga Mena}
\email{omena@ific.uv.es}
\affiliation{IFIC, Universidad de Valencia-CSIC, 46071, Valencia, Spain}

\author{Supriya Pan}
\email{supriya.maths@presiuniv.ac.in}
\affiliation{Department of Mathematics, Presidency University, 86/1 College Street,  Kolkata 700073, India }

%-----------------------------------------------------

\begin{abstract}
We investigate generalized interacting dark matter-dark energy scenarios with a time-dependent coupling parameter, allowing also for freedom in the neutrino sector. The models are tested in the phantom and quintessence regimes, characterized by an equation of state $w_x<-1$ and $w_x>-1$, respectively. Our analyses show that for some of the scenarios the existing tensions on the Hubble constant $H_0$ and on the clustering parameter $S_8$ can be significantly alleviated. The relief is either due to \textit{(a)} a dark energy component which lies within the phantom region; or  \textit{(b)} the presence of a dynamical coupling in quintessence scenarios. The inclusion of massive neutrinos into the interaction schemes does not affect neither the constraints on the cosmological parameters nor the bounds on the total number or relativistic degrees of freedom $N_{\rm eff}$, which are found to be extremely robust and, in general, strongly consistent with the canonical prediction $N_{\rm eff}=3.045$. The most stringent bound on the total neutrino mass $M_{\nu}$ is $M_{\nu}<0.116$~eV and it is obtained within a quintessence scenario in which the matter mass-energy density is only mildly affected by the presence of a dynamical dark sector coupling.
\end{abstract}


%--------------------------------------------------------
\maketitle
%-------------------------------------------------------

\section{Introduction}

Cosmological models where a non-gravitational interaction between the dark fluids of the universe, dark matter and dark energy, are still a very appealing and  interesting solution to the so-called \emph{why now?} problem. Early models were based on coupled quintessence scenarios~\cite{Carroll:1998zi,Wetterich:1994bg,Amendola-ide1,Amendola-ide2,Pavon:2005yx,delCampo:2008sr,delCampo:2008jx}, while more recent phenomenological approaches have adopted a number of possible parametrizations of the  energy exchange rate, see e.g.~\cite{Billyard:2000bh,Barrow:2006hia,Amendola:2006dg,He:2008tn,Valiviita:2008iv,Gavela:2009cy,Majerotto:2009np,Gavela:2010tm,Clemson:2011an,Pan:2012ki,Pan:2013rha,Yang:2014vza,Yang:2014gza,Nunes:2014qoa,vandeBruck:2015ida,Nunes:2016dlj,Kumar:2016zpg,Pan:2016ngu,vandeBruck:2016hpz,Mukherjee:2016shl,Sharov:2017iue,Kumar:2017dnp,Yang:2017yme,Yang:2017zjs,Mifsud:2017fsy,vandeBruck:2017idm,Kumar:2017bpv,Yang:2017ccc,Pan:2017ent,Yang:2018pej,Yang:2018ubt,Yang:2018xlt,Yang:2018qec,Martinelli:2019dau,Paliathanasis:2019hbi,Pan:2019jqh,Kumar:2019wfs,Yang:2019bpr,Yang:2019vni,Barrow:2019jlm,Pan:2019gop,Papagiannopoulos:2019kar,Yang:2020zuk,Pan:2020zza,Pan:2020mst,Pan:2020bur,Lucca:2020zjb}. Following our pioneering previous work~\cite{Yang:2019uzo} we shall consider here a time-dependent coupling in non-minimal cosmologies. Given the fact that neutrinos can play a non-standard role within non-minimal dark energy scenarios~\cite{Giusarma:2016phn,Gerbino:2016sgw,Vagnozzi:2017ovm,Vagnozzi:2018jhn,Giusarma:2018jei,Choudhury:2018byy,Vagnozzi:2018pwo,Vagnozzi:2019utt,Yang:2019uog,Yang:2020uga,Hagstotz:2020ukm}, we extend our previous analyses by inspecting the impact of neutrino properties within  interacting cosmologies  with a time-dependent coupling. We also generalize the work of Ref.~\cite{Yang:2019uzo} with the inclusion of a constant dark energy state parameter that may freely vary in a certain region. This picture also entails the case of a coupling parameter that remains constant in cosmic time. For our analyses we have assumed that our  universe is homogeneous and isotropic, that is, its geometry is well described by the Friedmann-Lema\^{i}tre-Robertson-Walker line element. In order to perform robust statistical analyses, we shall make use of various cosmological datasets such as the Cosmic Microwave Background radiation, Baryon Acoustic Oscillation distance measurements, and, finally, a local measurement of the Hubble constant from the Hubble Space Telescope.

The manuscript has been organized as follows: In Sec. ~\ref{sec-efe} we briefly introduce the gravitational equations for the two interacting dark fluids. Section~\ref{sec-data} describes the observational data, methodology and the priors imposed on the cosmological parameters. Section~\ref{sec-results} presents the current observational constraints on the interacting cosmic scenarios considered here. Section~\ref{sec-conclu} contains our main conclusions.

\section{Interacting dark sectors: Gravitational equations}
\label{sec-efe}
Observations suggest that at large scales, our universe is homogeneous and isotropic and therefore well described by the Friedmann-Lema\^{i}tre-Robertson-Walker (FLRW) line element 

\begin{eqnarray}
ds^2 =  - dt^2 + a^2 (t) \left[\frac{dr^2}{1-\kappa r^2} + r^2 \left(d\theta ^2 + \sin^2 \theta d \phi^2 \right) \right]~,  
\end{eqnarray}
where $a(t)$ is the expansion scale factor of the universe and $(t, r, \theta, \phi)$ are the co-moving coordinates. Having specified the metric of the underlying geometry of our universe, we assume in the following that the gravitational sector of the universe is described by General Relativity, the matter sector is minimally coupled to gravity, and, finally, that there is a non-gravitational interaction between the dark sectors of the universe, namely, between the pressureless dark matter (DM) and the dark energy (DE) fluids:

\begin{eqnarray}
&&\dot{\rho}_c + 3 H \rho_c = - Q~,\\
&&\dot{\rho}_x + 3 H (1+w_x) \rho  =  Q~,
\end{eqnarray}
where $H \equiv \dot{a}/a$ is the Hubble rate of the FLRW universe; $\rho_c$ ($p_c $), $\rho_x$ ($p_x$) are the energy density (pressure) for DM and DE respectively (albeit the DM fluid, being pressureless here, has $p_c  = 0$), $w_x  = p_x/\rho_x$ denotes the barotropic DE equation of state parameter (assumed here to be constant) and, finally, $Q$ determines the interaction rate between DM and DE. In general, when a specific form of the interaction rate is given, one can solve either analytically or numerically the background evolution for $\rho_c$ and $\rho_x$. We shall explore here the (time-dependent) interaction models of Ref.~\cite{Yang:2019uzo}:
\begin{eqnarray}
&\rm{IDE1}:& Q = 3 \xi (a) H \rho_{x}, \label{ide1}\\
&\rm{IDE2}:& Q= 3 \xi (a) H \frac{\rho_c \rho_{x}}{\rho_c+ \rho_{x}},\label{ide2}
\end{eqnarray}
where $\xi (a)$ is a 
time-dependent dimensionless coupling parameter. Similar to our earlier work~\cite{Yang:2019uzo}, we keep the parametrization of $\xi (a)$ as follows 
\begin{eqnarray}
\xi (a) = \xi_0 + \xi_a \; (1-a)~, \label{tot_xi}
\end{eqnarray}
where $\xi_0$ and $\xi_a$ are real constants. 
Finally, based on the stability criteria of the perturbation evolution~\cite{Valiviita:2008iv,Gavela:2009cy}, we shall classify the models as

\begin{eqnarray}
{\rm IDE1p:}\;\; w_x< -1,\; \xi_0 < 0,\; \xi_a < 0, \label{1p}\\
{\rm IDE1q:}\;\; w_x> -1,\; \xi_0 > 0,\; \xi_a > 0, \label{1q}
\end{eqnarray}
\noindent for the IDE1 case, and, equivalently,
  
  \begin{eqnarray}
{\rm IDE2p:}\;\; w_x< -1,\; \xi_0 < 0,\; \xi_a < 0, \label{2p}\\
{\rm IDE2q:}\;\; w_x> -1,\; \xi_0 > 0,\; \xi_a > 0, \label{2q}
\end{eqnarray}

\noindent for the IDE2 model, where {\rm p} and {\rm q} in {\rm IDEp} and {\rm IDEq} stand for \emph{phantom} and \emph{quintessence} regimes, respectively.

\section{Observational data and methodology}
\label{sec-data}

In the following we briefly describe the cosmological data sets used in this work. 

\begin{itemize}
    \item {\bf Cosmic Microwave Background (CMB)}: our  default data set is the one containing the latest CMB temperature and polarization measurements  in both the high and low multipole regions, i.e. {\it Plik TT,TE,EE + lowl + lowE}, from the final 2018 Planck legacy release~\cite{Aghanim:2018eyx,Aghanim:2018oex,Aghanim:2019ame}.
    
    \item {\bf Baryon Acoustic Oscillations (BAO)}: we make use of several BAO measurements from different cosmological observations, as considered by the Planck collaboration~\cite{Aghanim:2018eyx}: 6dFGS~\cite{Beutler:2011hx}, SDSS-MGS~\cite{Ross:2014qpa}, and BOSS DR12~\cite{Alam:2016hwk} surveys.

    \item {\bf Hubble constant Gaussian prior (R19)}: we assume a Gaussian prior on the Hubble constant, in agreement with that obtained by the SH0ES collaboration in 2019, i.e. $H_0 = 74.03 \pm 1.42$ km/s/Mpc at $68\%$ CL~\cite{Riess:2019cxk}.

\end{itemize}

For the analysis of the cosmological data, we adopt a fiducial model described by nine cosmological parameters. In particular, we vary the six parameters of the standard $\Lambda$CDM model, i.e. the baryon energy density $\Omega_{\rm b}h^2$, the cold dark matter energy density $\Omega_{\rm c}h^2$, the ratio between the sound horizon and the angular diameter distance at decoupling $100\theta_{MC}$, the reionization optical depth 
$\tau$, the spectral index $n_{s}$ and the amplitude of the scalar primordial power spectrum $A_{s}$. In addition we vary the three parameters of the dark sector physics considered here, i.e. the DE equation of state $w_{x}$ and the strength of the coupling, parametrized by $\xi_0$ and $\xi_a$, see Eq.~(\ref{tot_xi}). The parameter space will therefore be described by: 

\begin{eqnarray}
\mathcal{P} \equiv\Bigl\{\Omega_{b}h^2, \Omega_{c}h^2, 100\theta_{MC}, \tau, n_{s}, log[10^{10}A_{s}], \notag \\ 
\xi_0, \xi_a, w_{x} \Bigr\}~.
\label{eq:parameter_space1}
\end{eqnarray}
As aforementioned, the stability of the perturbation evolution restricts the IDE scenarios to two phantom cases ($w_{x}<-1$) ({\rm IDE1p}, Eq.(\ref{1p}) and {\rm IDE2p}, Eq.(\ref{2p})) with $\xi_0<0$ and $\xi_a<0$ and two quintessence regimes ($w_{x}>-1$) ({\rm IDE1q}, Eq.(\ref{1q}) and {\rm IDE2q}, Eq.(\ref{2q})) with $\xi_0>0$ and $\xi_a>0$. Table~\ref{tab:priors} lists the priors on all the parameters considered in this work. 

We shall also consider an enlarged cosmological scenario with eleven parameters, allowing the sum of the neutrino masses $M_{\nu}$ and the number or relativistic degrees of freedom $N_{\rm eff}$ to freely vary (IDE $+$ $M_{\nu}$ $+$ $N_{\rm eff}$): 

\begin{eqnarray}
\mathcal{P} \equiv\Bigl\{\Omega_{b}h^2, \Omega_{c}h^2, 100\theta_{MC}, \tau, n_{s}, \log[10^{10}A_{s}], \nonumber\\
\xi_0, \xi_a, w_x, M_{\nu}, N_{\rm eff}\Bigr\}~,
\label{eq:parameter_space2}
\end{eqnarray}

\noindent and also in this case we will have four cases, depending on the scenario of IDE considered and on the phantom or quintessence regime, i.e. {\rm IDE1p}, {\rm IDE1q}, {\rm IDE2p} and {\rm IDE2q}, respectively.

To derive the constraints on the cosmological parameters we shall use a modified version with models IDE1 and IDE2 implemented of the publicly available Markov Chain Monte Carlo code \texttt{CosmoMC}~\cite{Lewis:2002ah,Lewis:1999bs} package. This version supports the new 2018 Planck likelihood~\cite{Aghanim:2019ame} and uses a convergence diagnostic following the Gelman-Rubin criteria~\cite{Gelman-Rubin}.

\begin{table}
\begin{center}
\renewcommand{\arraystretch}{1.4}
\begin{tabular}{|c@{\hspace{1 cm}}|@{\hspace{0.5 cm}} c@{\hspace{0.5 cm}}|@{\hspace{1 cm}} c|}
\hline
\textbf{Parameter}                    & \textbf{Prior}& \textbf{Prior}  \\
\hline\hline
$\Omega_{b} h^2$             & $[0.005,0.1]$& $[0.005,0.1]$\\
$\Omega_{c} h^2$             & $[0.01,0.99]$& $[0.01,0.99]$\\
$\tau$                       & $[0.01,0.8]$& $[0.01,0.8]$\\
$n_s$                        & $[0.5, 1.5]$& $[0.5, 1.5]$\\
$\log[10^{10}A_{s}]$         & $[2.4,4]$& $[2.4,4]$\\
$100\theta_{MC}$             & $[0.5,10]$& $[0.5,10]$\\
$w_x$                        & $[-3, -1]$& $[-1, 0]$ \\
$\xi_0$                      & $[-1, 0]$& $[0, 1]$\\
$\xi_a$                      & $[-1, 0]$& $[0,1]$\\ 
$M_{\nu}$                    & $[0,1]$ & $[0,1]$ \\
$N_{\rm eff}$                & $[0.05,10]$& $[0.05,10]$  \\
\hline
\end{tabular}
\end{center}
\caption{The table shows the flat priors imposed on various free parameters of the cosmological scenarios to be discussed in this work. }
\label{tab:priors}
\end{table}

\section{Results}
\label{sec-results}

\subsection{{\rm IDE1}}
In the following we shall show the results obtained for the IDE1 scenario presented in Eq.~(\ref{ide1}), both in the phantom and in the quintessence regimes, and with and without varying the neutrino sector.

\subsubsection{{\rm IDE1p}}

\begin{figure*}
\includegraphics[width=0.55\textwidth]{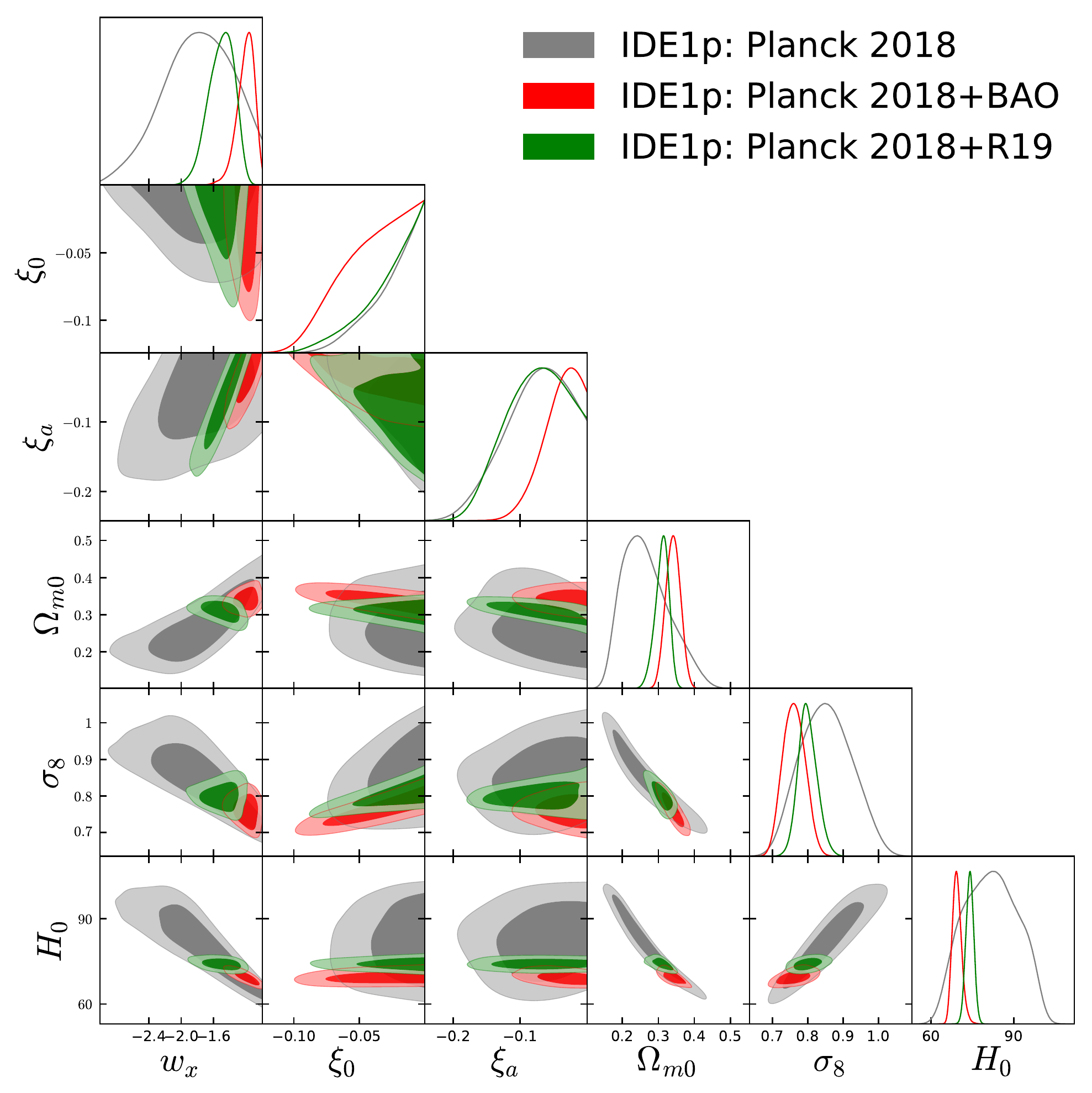}
\caption{One-dimensional marginalized posterior distributions and $68\%$ and $95\%$ CL two-dimensional contours for the interacting scenario {\rm IDE1p} for the  cosmological dataset combinations  considered  in this study. }
\label{fig:IDE1p}
\end{figure*}

The results for the IDE1 model in the phantom regime, i.e. with $w_x<-1$, $\xi_0 < 0$ and $\xi_a < 0$ are reported in Tab.~ \ref{tab:resultsIDE1p} and Fig.~\ref{fig:IDE1p}.

For an interacting dark energy with a phantom-like equation of state, the CDM energy density $\Omega_ch^2$ is larger than in the $\Lambda$CDM model, provided the energy transfer is from the DE to the DM sector~\cite{DiValentino:2019jae,Yang:2020uga}. Furthermore, due  to the strong degeneracy between $w_x$ and $H_0$, see Fig.~\ref{fig:IDE1p}, the Hubble constant is almost unconstrained for CMB only data. The well known $H_0$ tension is strongly alleviated within this model. While $\xi_0$ has only lower limit for all the combinations of data  considered here, being therefore consistent with a vanishing interaction at present, we find $\xi_a$ different from zero at one standard deviation for the CMB only ($\xi_a=-0.077^{+0.064}_{-0.032}$ at 68\% CL) and for the CMB+R19 ($\xi_a=-0.077^{+0.059}_{-0.037}$ at 68\% CL) cases. A very interesting  feature of this model is the strong evidence for a phantom-like equation of state $w_x<-1$  for all the data combinations, with a statistical significance increasing from 1$\sigma$ for the CMB only case ($w_x=-1.80^{+0.49}_{-0.39}$ at 68\% CL), to about $2\sigma$ for CMB+BAO.
Finally, the $S_8$ parameter moves towards lower values for the CMB only case, enough to bring it in agreement with the cosmic shear experiments DES~\cite{Abbott:2017wau,Troxel:2017xyo}, KiDS-450~\cite{Kuijken:2015vca,Hildebrandt:2016iqg,Conti:2016gav}, CFHTLenS~\cite{Heymans:2012gg,  Erben:2012zw,Joudaki:2016mvz}, or the combination of KiDS+VIKING-450 and DES-Y1~\cite{Asgari:2019fkq} , i.e.  $S_8=0.789\pm0.037$ at 68\% CL. However, when the BAO or the R19 priors are added to the CMB, the $S_8$ values are increased, restoring the tension at more than 3 standard deviations. 

Finally, in Table~\ref{tab:chi2}, we show the $\chi^2$ values for this model, as well as other models considered in this work, for all the observational datasets employed here. In the same Table~\ref{tab:chi2}, we have also shown the $\chi^2$ values for the non-interacting scenario $w$CDM model as the reference model.   
From Table~\ref{tab:chi2} we can see that the $\chi^2$ values obtained for this scenario (i.e., IDE1p)  are improved with respect to the $w$CDM model of about $2$ (for Planck 2018+BAO) and $4.5$ (Planck 2018+R19), even if in our case we have two more degrees of freedom compared to the $w$CDM model.

\begingroup                                         %\squeezetable    
\begin{center}
\begin{table*} 
\begin{tabular}{cccccccccccccc} 
\hline\hline 
Parameters & Planck 2018 & Planck 2018+BAO & Planck 2018+R19 \\ \hline
$\Omega_c h^2$ & $    0.148_{-    0.019}^{+    0.017}$ & $    0.141_{-    0.014}^{+    0.013}$ & $    0.147_{-    0.018}^{+    0.014}$   \\

$\Omega_b h^2$ & $    0.02246_{-    0.00030}^{+    0.00029}$ 
& $    0.02246_{-    0.00030}^{+    0.00030}$  & $    0.02244_{-    0.00030}^{+    0.00031}$  \\

$100\theta_{MC}$ & $    1.0395_{-    0.0011}^{+    0.0012}$ & $    1.03989_{-    0.00084}^{+    0.00093}$ & $    1.0396_{-    0.0010}^{+    0.0010}$  \\

$\tau$ & $    0.053_{-    0.015}^{+    0.016}$ & $    0.055_{-    0.015}^{+    0.016}$  & $    0.053_{-    0.015}^{+    0.015}$  \\

$n_s$ & $    0.9671_{-    0.0091}^{+    0.0088}$ & $    0.9678_{-    0.0088}^{+    0.0087}$  & $    0.9668_{-    0.0090}^{+    0.0087}$  \\

${\rm{ln}}(10^{10} A_s)$ & $    3.039_{-    0.030}^{+    0.031}$ & $    3.043_{-    0.031}^{+    0.032}$  & $    3.040_{-    0.031}^{+    0.031}$ \\

$w_x$ & $>-2.53$ & $   -1.21_{-    0.21}^{+    0.20}$  & $   -1.50_{-    0.31}^{+    0.30}$ \\

$\xi_0$ & $>-0.061$ & $>-0.084$  & $>-0.071$  \\

$\xi_a$ & $ >-0.16$ & $>-0.091$  & $ >-0.15$  \\

$\Omega_{m0}$ & $    0.27_{-    0.11}^{+    0.13}$ & $    0.341_{-    0.039}^{+    0.039}$  & $    0.310_{-    0.039}^{+    0.036}$  \\

$\sigma_8$ & $    0.85_{-    0.14}^{+    0.14}$ &  $    0.761_{-    0.058}^{+    0.061}$ & $    0.800_{-        0.049}^{+        0.055}$  \\

$H_0{\rm [km/s/Mpc]}$ & $ >63.9$ & $   69.4_{-      3.3}^{+      3.4}$  & $   74.0_{-      2.7}^{+      2.7}$ \\

$S_8$ & $    0.789_{-        0.068}^{+       0.067}$ &  $    0.810_{-       0.035}^{+       0.033}$ & $    0.812_{-        0.042}^{+      0.037}$ \\

\hline\hline
\end{tabular} 
\caption{95\% CL constraints on the interacting scenario {\rm IDE1p} using CMB from Planck 2018, BAO and local measurements of $H_0$ from R19. }
\label{tab:resultsIDE1p}
\end{table*}                                         \end{center}                                        
\endgroup   

\subsubsection{{\rm IDE1p} $+$ $M_{\nu}$ $+$ $N_{\rm eff}$}

\begin{figure*}
\includegraphics[width=0.66\textwidth]{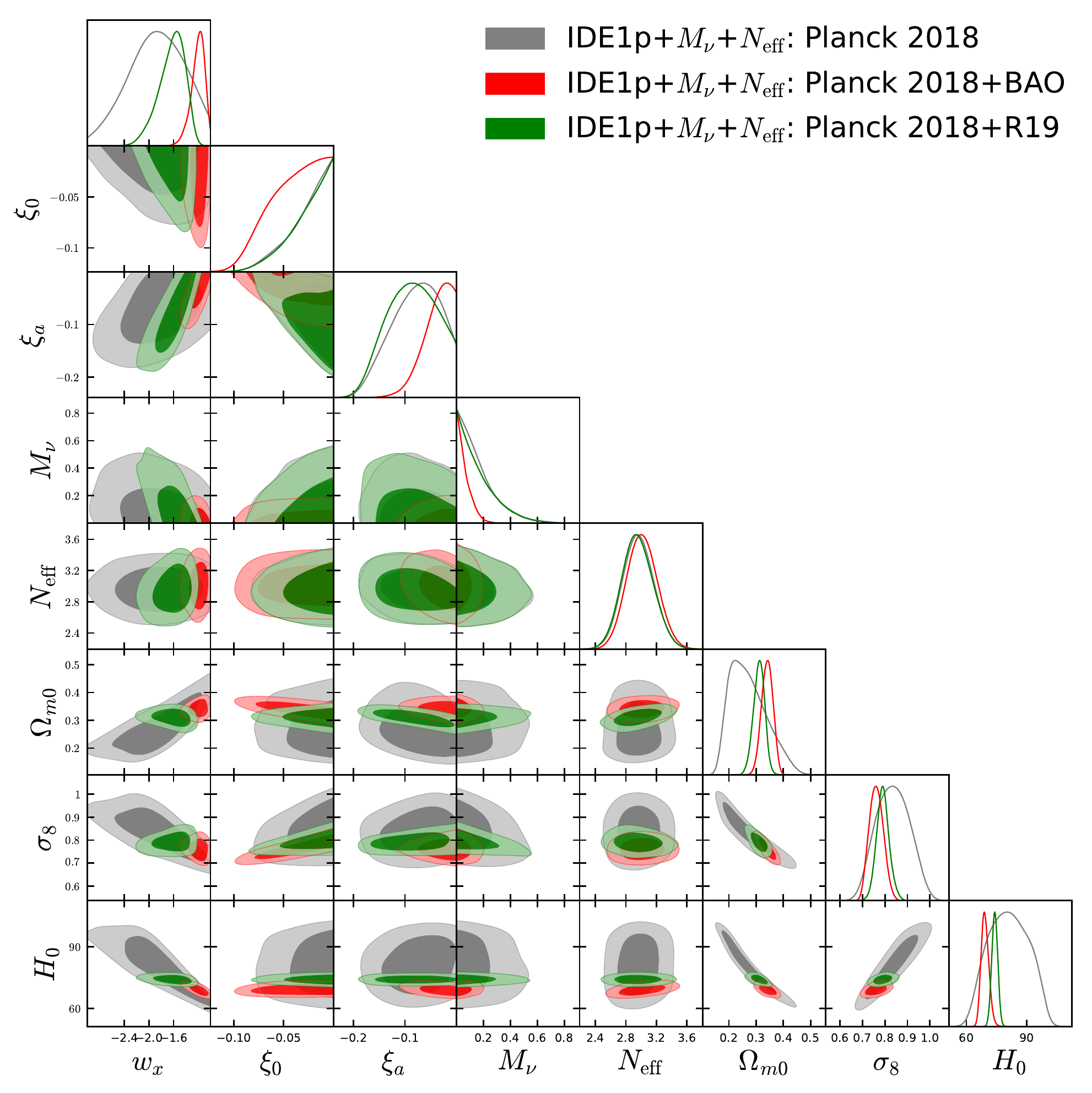}
\caption{One-dimensional marginalized posterior distributions and $68\%$ and $95\%$ CL two-dimensional contours for the interacting scenario {\rm IDE1p} $+$ $M_{\nu}$ $+$ $N_{\rm eff}$ for the  cosmological dataset combinations  considered  in this study.}
\label{fig:IDE1pnu}
\end{figure*}

The results for the IDE1 model in the phantom regime with the addition of the neutrino parameters, i.e. $M_{\nu}$ and $N_{\rm eff}$, are shown in Tab.~\ref{tab:resultsIDE1pnu} and Fig.~\ref{fig:IDE1pnu}.

The constraints from the previous section on the cosmological parameters and their correlations ({\rm IDE1p}) are barely affected by allowing $M_\nu$ and $N_{\rm eff}$ to freely vary simultaneously.
In particular, $\Omega_ch^2$ is larger than in $\Lambda$CDM model and the Hubble constant tension with R19 is solved within 3$\sigma$ even when BAO data are included.
Also in this case $\xi_0$ has just a lower limit and is consistent with zero, while $\xi_a$ is different from zero at one standard deviation for the CMB only ($\xi_a=-0.081^{+0.060}_{-0.037}$ at 68\% CL) and CMB+R19 ($\xi_a=-0.087^{+0.055}_{-0.048}$ at 68\% CL) cases, but consistent with zero when BAO data are included. 

The indication for a phantom equation of state $w_x<-1$ is instead present for all the dataset combinations with a statistical significance always larger two standard  deviations, even for the CMB only case.
The neutrino sector parameters $M_\nu$ and $N_{\rm eff}$ are mostly uncorrelated with the other cosmological parameters, with the exception of $w_x$ that strongly anti-correlates with the total neutrino mass,  $M_{\nu}$. The existence of anti-correlation between $w_x$ and $M_{\nu}$ is not new, in fact, in the usual non-interacting $w (z)$CDM cosmology, this has been already pointed out  \cite{Vagnozzi:2018jhn}, however, the interesting observation in this case that we find, even if the presence scenario allows an interaction in the dark sector, 
this anti-correlation does not get affected due to such interaction. The preference for $w_x<-1$ is therefore the reason for the much weaker upper limits on $M_\nu$ with respect to the same combinations of data within a $\Lambda$CDM model~\cite{Vagnozzi:2018jhn}. 
The most stringent limit we find on the sum of the neutrino masses is when adding BAO data to the CMB, i.e. $M_\nu<0.162$ eV at 95\% CL. 

Regarding the constraints on the effective number of relativistic degrees of freedom $N_{\rm eff}$, these are completely unaffected by the inclusion of the interaction $\xi(a)$: in this scenario $N_{\rm eff}$ is always consistent with its expected value of  $3.045$~\cite{Mangano:2005cc,deSalas:2016ztq}.

In Table~\ref{tab:chi2} we can see that the $\chi^2$ values for this scenario (i.e., IDE1p $+$ $M_{\nu}$ $+$ $N_{\rm eff}$) are always below compared to the $w$CDM $+$ $M_{\nu}$ $+$ $N_{\rm eff}$ model, up to $4.2$ for Planck 2018+R19. We note that the model IDE1p $+$ $M_{\nu}$ $+$ $N_{\rm eff}$ has two more degrees of freedom compared to the $w$CDM $+$ $M_{\nu}$ $+$ $N_{\rm eff}$ model.

\begingroup                                          %\squeezetable                                     
\begin{center}                                       \begin{table*}                                       \begin{tabular}{cccccccccccccccc}                    \hline\hline 
Parameters & Planck 2018 & Planck 2018+BAO & Planck 2018+R19\\ \hline

$\Omega_c h^2$ & $    0.147_{-        0.019}^{+     0.018}$ & $    0.140_{-       0.017}^{+    0.017}$ & $    0.146_{-       0.019}^{+        0.017}$ \\

$\Omega_b h^2$ & $    0.02235_{-        0.00049}^{+      0.00050}$ & $    0.02243_{-        0.00044}^{+       0.00046}$ & $    0.02234_{-    0.00048}^{+        0.00047}$ \\

$100\theta_{MC}$ & $    1.0396_{-        0.0012}^{+        0.0013}$ & $    1.0400_{-        0.0012}^{+        0.0012}$ & $    1.0396_{-        0.0012}^{+        0.0013}$ \\

$\tau$ & $    0.053_{-        0.015}^{+        0.015}$ &  $    0.055_{-       0.015}^{+       0.016}$  & $    0.053_{-       0.015}^{+    0   0.016}$ \\

$n_s$ & $    0.963_{-        0.018}^{+       0.018}$ &  $    0.966_{-        0.017}^{+      0.018}$   & $    0.963_{-        0.018}^{+       0.018}$  \\

${\rm{ln}}(10^{10} A_s)$ & $    3.036_{-        0.037}^{+        0.036}$ &  $    3.041_{-       0.037}^{+        0.037}$ & $    3.036_{-        0.035}^{+       0.037}$  \\

$w_x$ & $   -1.88_{-        0.81}^{+       0.83}$ & $   -1.21_{-       0.22}^{+        0.20}$  & $   -1.63_{-        0.44}^{+        0.39}$ \\

$\xi_0$ & $>-0.067$ & $>-0.083$  & $>-0.066$ \\

$\xi_a$ & $ >-0.16$ &  $>-0.090$  & $  >-0.17$ \\

$\Omega_{m0}$ & $    0.27_{-       0.11}^{+       0.14}$ & $    0.341_{-     0.041}^{+       0.041}$ & $    0.311_{-       0.042}^{+      0.039}$  \\

$\sigma_8$ & $    0.84_{-        0.13}^{+     0.14}$ &  $    0.762_{-        0.060}^{+     0.061}$  & $    0.791_{-      0.054}^{+       0.059}$ \\

$H_0 {\rm [km/s/Mpc]}$ & $   81_{-     17}^{+     18}$ &  $   69.2_{-       3.6}^{+       3.8}$  & $   74.0_{-       2.9}^{+     2.8}$ \\

$M_\nu {\rm [eV]}$ & $<0.438$ & $<0.162$  & $ <0.437$ \\

$N_{\rm eff}$ & $    2.96_{-        0.38}^{+        0.40}$  & $    3.01_{-        0.38}^{+       0.40}$  & $    2.96_{-       0.39}^{+        0.41}$ \\

$\Omega_\nu h^2$ & $<0.0047$ &  $<0.0017$ & $ <0.0046$  \\

$S_8$ & $    0.781_{-        0.076}^{+       0.071}$ & $    0.811_{-        0.037}^{+      0.037}$  & $    0.803_{-       0.048}^{+      0.046}$ \\

\hline\hline                                         \end{tabular}                                        
\caption{95\% CL constraints on the interacting scenario {\rm IDE1p} $+$ $M_{\nu}$ $+$ $N_{\rm eff}$ using CMB from Planck 2018, BAO and local measurements of $H_0$ from R19. } \label{tab:resultsIDE1pnu}                           \end{table*}                                         \end{center}                                         
\endgroup

\subsubsection{{\rm IDE1q}}

\begin{figure*}
\includegraphics[width=0.55\textwidth]{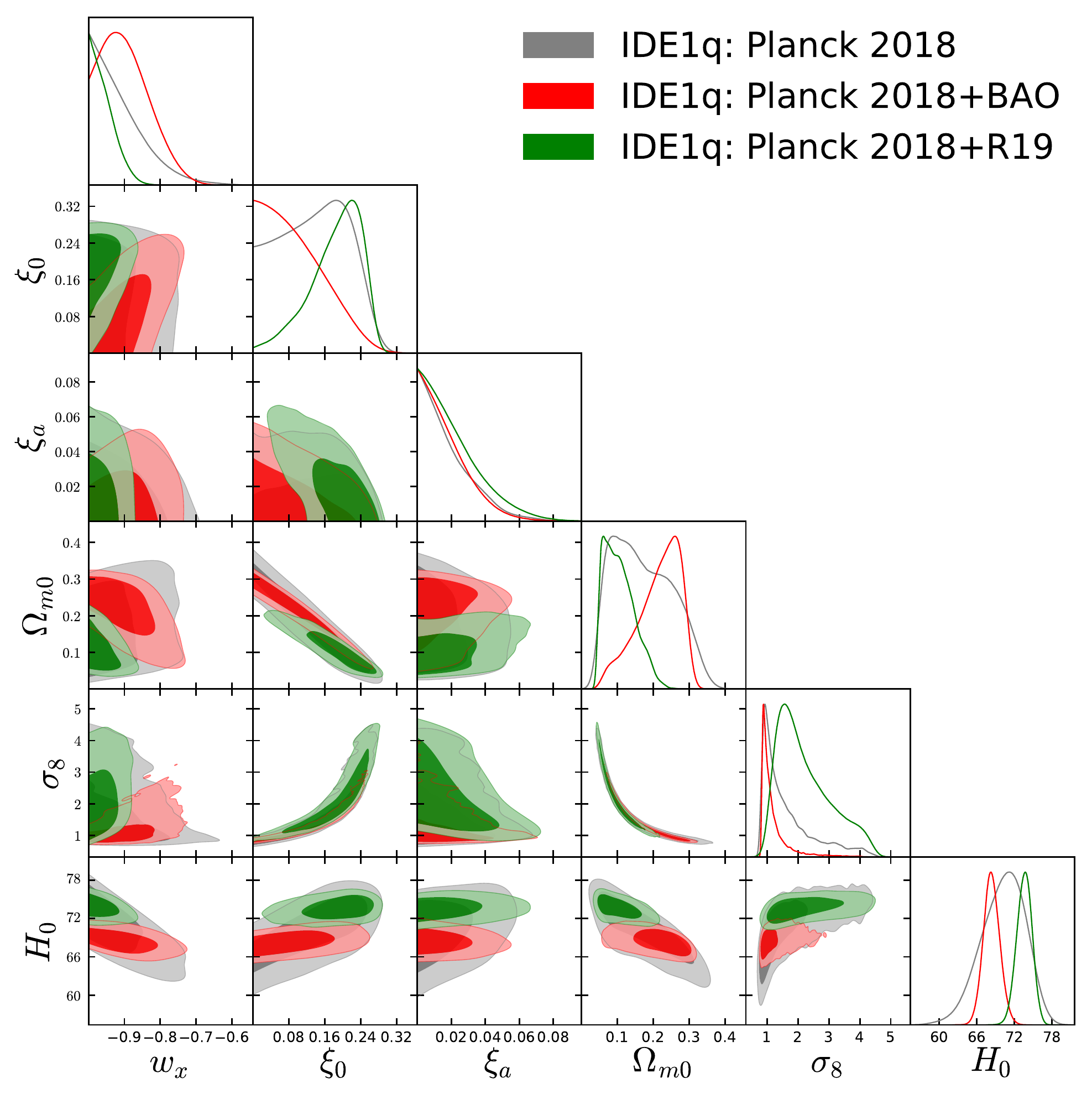}
\caption{One-dimensional marginalized posterior distributions and $68\%$ and $95\%$ CL two-dimensional contours for the interacting scenario {\rm IDE1q} for the  cosmological dataset combinations  considered  in this study. }
\label{fig:IDE1q}
\end{figure*}

The results for the IDE1 model in the quintessence regime, Eq.~(\ref{1q}), are reported in Tab.~\ref{tab:resultsIDE1q} and Fig.~\ref{fig:IDE1q}.

For an interacting dark energy with a quintessence-like equation of state, the CDM energy density $\Omega_ch^2$ is always smaller than in a $\Lambda$CDM model: indeed,  only an upper limit for this cosmological parameter is found~
\cite{DiValentino:2019ffd,DiValentino:2019jae,Yang:2020uga}.
The most interesting feature of this {\rm IDE1q} scenario is that, even if the well-known anti-correlation between $w_x$ and $H_0$ is present, see Fig.~\ref{fig:IDE1q}, the positive correlation between $\xi_0$ and $H_0$ shifts the Hubble constant towards higher values, solving the $H_0$ tension within 1$\sigma$ for the CMB only case ($H_0=70.2^{+4.1}_{-3.1}$ km/s/Mpc at 68\% CL).

Contrarily to the {\rm IDE1p} case, in this {\rm IDE1q} scenario the value of $\xi_0$, i.e. the interaction today, is found to be different from zero at low (high) significance for the CMB (CMB+R19) data. While for the CMB only and  the CMB+R19 cases only an upper limit on $w_x$  is found, an indication at 1$\sigma$ for $w_x>-1$ appears for CMB+BAO ($w_x=-0.895^{+0.040}_{-0.093}$ at 68\% CL). In this scenario, the $S_8$ parameter moves towards larger values, however the error bars are very large, enabling an agreement with cosmic shear experiments.

Finally, in Table~\ref{tab:chi2} we can see that the $\chi^2$ for this scenario (i.e., IDE1q) is systematically higher than the $w$CDM model, therefore it is disfavoured by the fit of the data.

\begingroup                                          %\squeezetable   
\begin{center}                                       \begin{table*}                                       \begin{tabular}{cccccccccccccc}                      \hline\hline                       
Parameters & Planck 2018 & Planck 2018+BAO & Planck 2018+R19 \\ \hline
$\Omega_c h^2$ & $<0.109$ & $    0.077_{-       0.058}^{+     0.043}$ & $<0.075$ \\

$\Omega_b h^2$ & $    0.02232_{-       0.00031}^{+      0.00029}$ & $    0.02233_{-      0.00029}^{+      0.00029}$  & $    0.02234_{-       0.00029}^{+        0.00029}$  \\

$100\theta_{MC}$ & $    1.0450_{-      0.0042}^{+     0.0048}$ & $    1.0436_{-       0.0030}^{+     0.0043}$  & $    1.0468_{-     0.0035}^{+    0.0034}$   \\

$\tau$ & $    0.054_{-       0.015}^{+      0.016}$ &  $    0.055_{-       0.015}^{+       0.016}$  & $    0.054_{-        0.015}^{+      0.016}$  \\

$n_s$ & $    0.9641_{-        0.0089}^{+      0.0088}$ &  $    0.9647_{-     0.0086}^{+        0.0082}$ & $    0.9645_{-        0.0086}^{+      0.0086}$   \\

${\rm{ln}}(10^{10} A_s)$ & $    3.046_{-       0.031}^{+      0.031}$ & $    3.046_{-    0.032}^{+     0.033}$  & $    3.045_{-    0.031}^{+        0.033}$  \\

$w_x$ & $<-0.77$ & $   <-0.77$  & $ <-0.89$  \\

$\xi_0$ & $ <0.25$ &  $<0.22$  & $    0.19_{-      0.12}^{+       0.10}$  \\

$\xi_a$ & $<0.046$ &  $<0.043$  & $ <0.054$  \\

$\Omega_{m0}$ & $    0.17_{-        0.14}^{+      0.16}$ & $    0.22_{-    0.13}^{+   0.10}$  & $    0.106_{-       0.071}^{+      0.086}$ \\

$\sigma_8$ & $    1.7_{-      1.2}^{+      2.0}$ &  $    1.2_{-    0.6}^{+       1.1}$  & $    2.2_{-    1.4}^{+       1.9}$  \\
$H_0 {\rm [km/s/Mpc]}$ & $   70.2_{-      7.1}^{+     6.7}$ &  $   68.4_{-      2.5}^{+      2.7}$  & $   73.6_{-    2.5}^{+      2.3}$  \\

$S_8$ & $    1.06_{-    0.31}^{+     0.49}$ & $    0.95_{-        0.19}^{+       0.33}$  & $    1.19_{-      0.38}^{+      0.45}$  \\

\hline\hline                                         \end{tabular}   
\caption{95\% CL constraints on the interacting scenario {\rm IDE1q} using CMB from Planck 2018, BAO and local measurements of $H_0$ from R19. }
\label{tab:resultsIDE1q}                             \end{table*}                                         \end{center}                                         
\endgroup

\subsubsection{{\rm IDE1q} $+$ $M_{\nu}$ $+$ $N_{\rm eff}$}

\begin{figure*}
\includegraphics[width=0.66\textwidth]{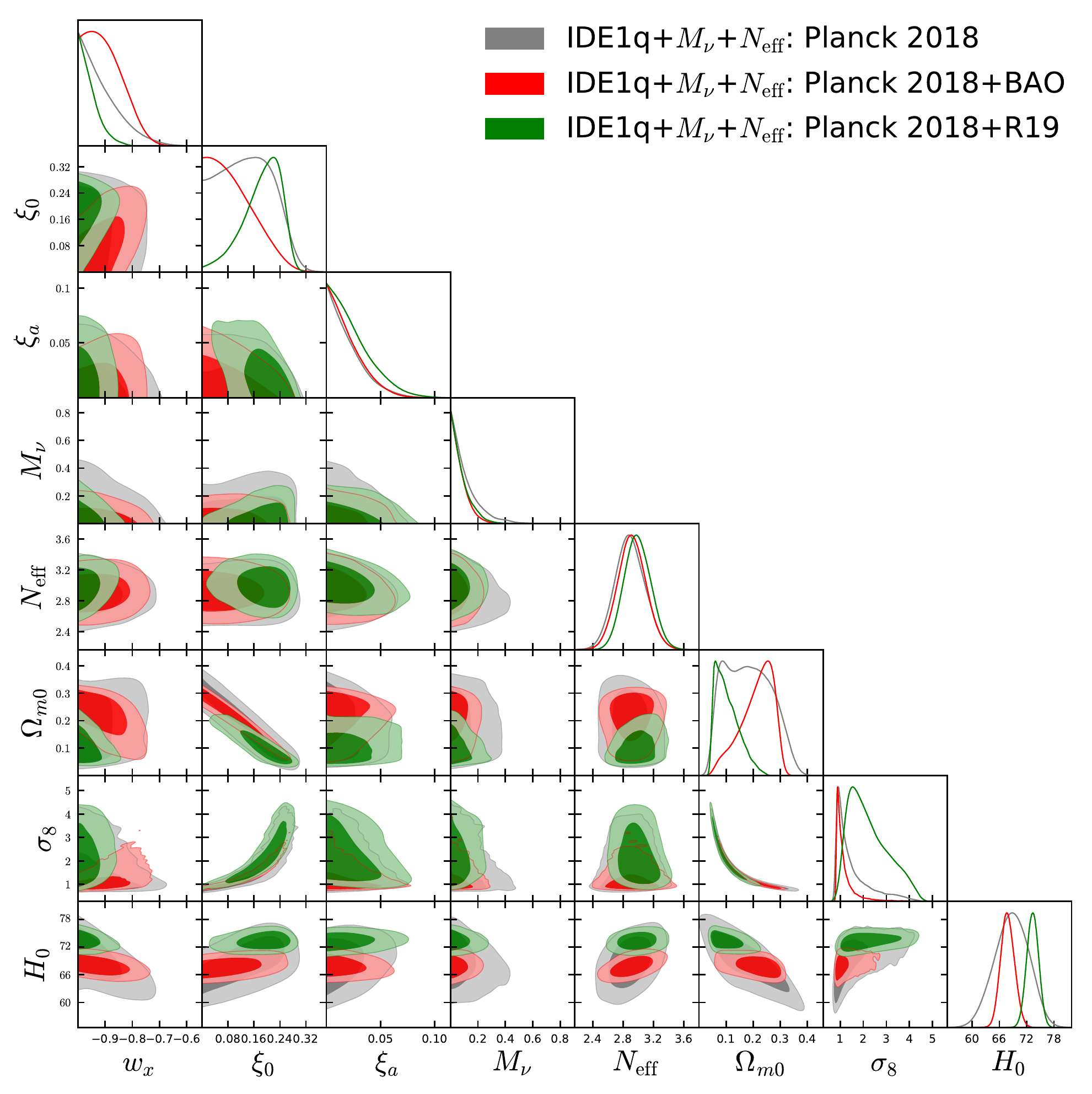}
\caption{One-dimensional marginalized posterior distributions and $68\%$ and $95\%$ CL two-dimensional contours for the interacting scenario {\rm IDE1q} $+$ $M_{\nu}$ $+$ $N_{\rm eff}$ for the  cosmological dataset combinations  considered  in this study. }
\label{fig:IDE1qnu}
\end{figure*}

The results for the IDE1 model in the quintessence regime  extended to include the neutrino parameters are shown in Tab.~\ref{tab:resultsIDE1qnu} and Fig.~\ref{fig:IDE1qnu}.

Similarly to the phantom case, both the constraints on the cosmological parameters and the correlations presented above are robust and are not affected by the introduction of the neutrino parameters $M_\nu$ and $N_{\rm eff}$. As in the previous section, $\xi_0$ is found to be different from zero at one standard deviation for the CMB only dataset ($\xi_0=0.137^{+0.087}_{-0.089}$ at 68\% CL), at several standard deviations for CMB+R19, and it has just an upper limit for the CMB+BAO case. In this extended scenario $\xi_a$ is always consistent with zero, as well as $w_x$ is consistent with $-1$ at $95\%$~CL for all the data combinations.

Also in this case the only important correlation between the neutrino sector and the remaining cosmological parameters is the one present between $M_\nu$ and $w_x$. However, in this quintessence regime, the CMB only upper limit on $M_\nu$ is stronger than the one found in the phantom regime (see Ref.~\cite{Vagnozzi:2018jhn}), and including the R19 prior this upper bound becomes even stronger ($M_\nu<0.221$ eV at 95\% CL). We note here that similar to the $w(z)$CDM case explored in  \cite{Vagnozzi:2018jhn} the anti-correlation between $M_\nu$ and $w_x$ remains unaltered in presence of the interaction between these dark sectors. This is an important point which clarifies that the anti-correlation between $M_\nu$ and $w_x$ seems to be independent of the coupling in the dark sector. The most stringent limit in this case we find on the sum of the neutrino masses is when adding BAO data to the CMB, i.e. $M_\nu<0.189$ eV at 95\% CL.

Finally, in this extended scenario (as in the phantom one), the constraints on the effective number of relativistic degrees of freedom $N_{\rm eff}$ are completely consistent with its canonical value $N_{\rm eff}=3.045$ for all the data combinations.

In Table~\ref{tab:chi2} we can see that the $\chi^2$ values for this scenario (i.e., IDE1q $+$ $M_{\nu}$ $+$ $N_{\rm eff}$) are always larger than the $w$CDM $+$ $M_{\nu}$ $+$ $N_{\rm eff}$ model. Therefore, this case is also disfavoured by the data.

\begingroup                                          %\squeezetable  
\begin{center}                                       \begin{table*}                                       \begin{tabular}{cccccccccccccc}                      \hline\hline                                                                                                                    
Parameters & Planck 2018 & Planck 2018+BAO & Planck 2018+R19 \\ \hline
$\Omega_c h^2$ & $ <0.108$ & $    0.075_{-       0.057}^{+       0.043}$ &  $<0.076$ \\

$\Omega_b h^2$ & $    0.02217_{-      0.00046}^{+       0.00046}$ & $    0.02223_{-     0.00041}^{+     0.00040}$ & $    0.02229_{-       0.00041}^{+     0.00041}$ \\

$100\theta_{MC}$ & $    1.0450_{-       0.0040}^{+     0.0048}$ &  $    1.0439_{-     0.0031}^{+        0.0043}$ & $    1.0470_{-      0.0039}^{+        0.0034}$  \\

$\tau$ & $    0.054_{-      0.015}^{+    0.016}$ & $    0.054_{-        0.015}^{+       0.015}$ &  $    0.054_{-       0.015}^{+      0.016}$ \\

$n_s$ & $    0.958_{-     0.017}^{+     0.018}$ & $    0.960_{-       0.016}^{+       0.016}$ &  $    0.963_{-        0.015}^{+      0.015}$ \\

${\rm{ln}}(10^{10} A_s)$ & $    3.038_{-       0.037}^{+       0.039}$ & $    3.039_{-      0.036}^{+        0.036}$ & $    3.043_{-      0.035}^{+      0.037}$   \\

$w_x$ & $<-0.105$ &  $<-0.781$ &  $<-0.881$ \\

$\xi_0$ & $  <0.26$ & $ <0.23$ &  $    0.19_{-       0.12}^{+      0.10}$ \\

$\xi_a$ & $ <0.052$ & $ <0.050$ &   $ <0.060$\\

$\Omega_{m0}$ & $    0.18_{-       0.14}^{+     0.15}$ & $    0.21_{-       0.13}^{+      0.10}$ & $    0.104_{-     0.070}^{+      0.089}$ \\

$\sigma_8$ & $    1.6_{-        1.0}^{+    1.9}$ & $    1.2_{-       0.6}^{+      1.1}$ & $    2.2_{-      1.4}^{+      1.8}$  \\

$H_0 {\rm [km/s/Mpc]}$ & $   68.7_{-      7.4}^{+       6.8}$ & $   67.8_{-      2.8}^{+       3.1}$ &  $   73.3_{-        2.5}^{+     2.5}$ \\

$M_\nu {\rm [eV]}$ & $<0.326$ & $<0.189$ &  $<0.221$ \\

$N_{\rm eff}$ & $    2.89_{-        0.37}^{+        0.39}$ & $    2.92_{-      0.36}^{+     0.37}$ & $    2.99_{-       0.33}^{+       0.35}$  \\

$\Omega_\nu h^2$ & $<0.0034$ & $<0.0020$ &  $<0.0024$ \\

$S_8$ & $    1.04_{-       0.28}^{+     0.47}$ & $    0.95_{-       0.19}^{+   0.32}$ & $    1.19_{-   0.38}^{+     0.43}$\\

\hline\hline                                         \end{tabular}                                        \caption{95\% CL constraints on the interacting scenario {\rm IDE1q} $+$ $M_{\nu}$ $+$ $N_{\rm eff}$ using CMB from Planck 2018, BAO and local measurements of $H_0$ from R19.  }
\label{tab:resultsIDE1qnu}                           \end{table*}                                         \end{center}                                         
\endgroup
\subsection{{\rm IDE2}}
In the  following we shall show the bounds on the cosmological parameters obtained for the {\rm IDE2} scenario, see Eq.~(\ref{ide2}), both in the phantom and in the quintessence regimes, and with and without varying the neutrino sector.

\subsubsection{{\rm IDE2p}}

\begin{figure*}
\includegraphics[width= 0.55\textwidth]{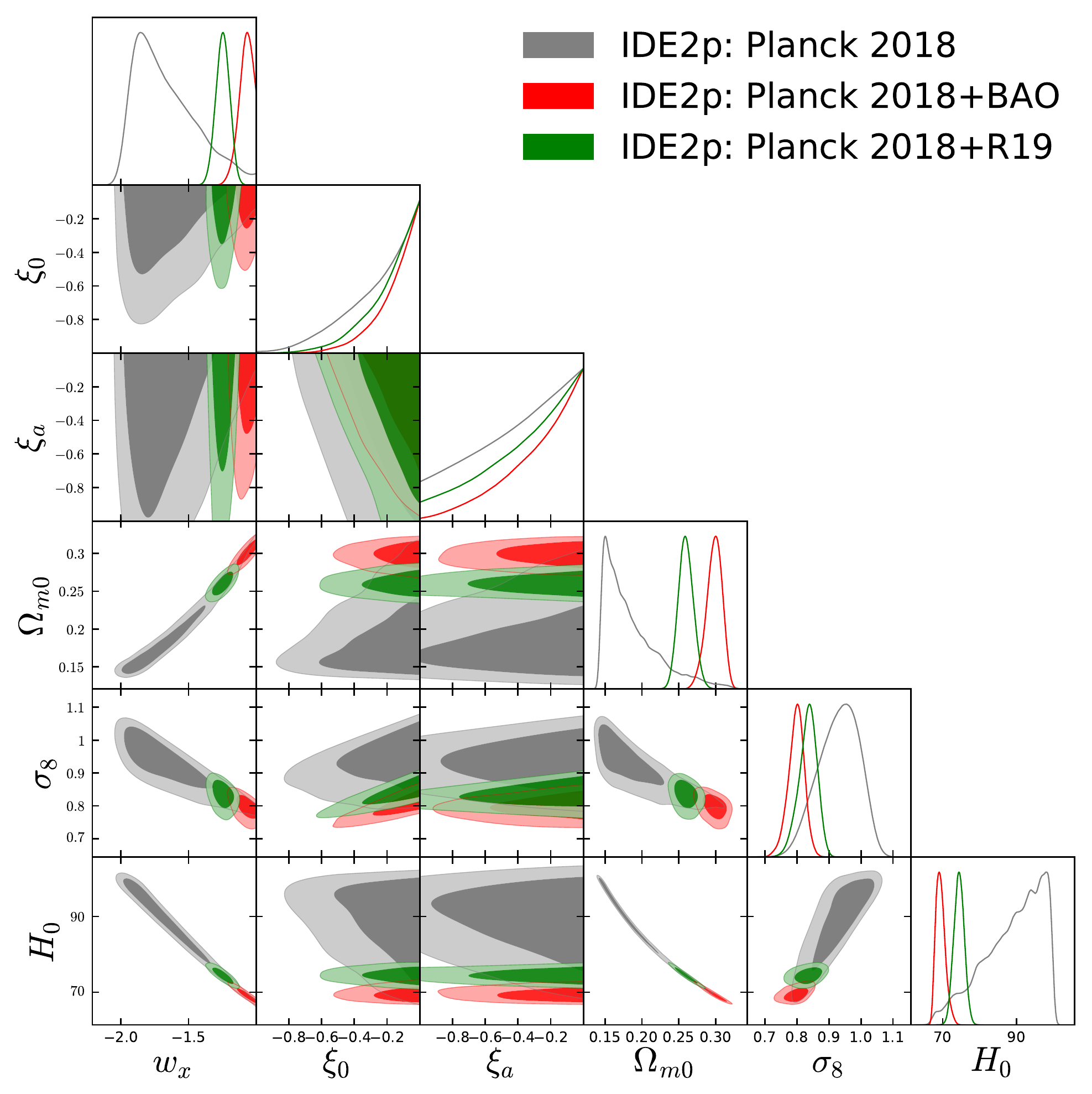}
\caption{One-dimensional marginalized posterior distributions and $68\%$ and $95\%$ CL two-dimensional contours for the interacting scenario {\rm IDE2p} for the  cosmological dataset combinations  considered  in this study.}
\label{fig:IDE2p}
\end{figure*}

The results for the {\rm IDE2} model in the phantom regime are reported in Tab.~\ref{tab:resultsIDE2p} and Fig.~\ref{fig:IDE2p}.

In the {\rm IDE2} model the interaction rate depends on both the cold dark matter density and the dark energy density. For this reason the flux of energy in the dark sector, from DE to DM and vice versa, can change with time. In this scenario, the bound on the cold dark matter energy density $\Omega_ch^2$ is in perfect agreement with that obtained within a $\Lambda$CDM model, as we can notice from Tab.~\ref{tab:resultsIDE2p}.
The well-known negative correlation present between $w_x$ and $H_0$ when $w_x$ is in the phantom regime (see Fig.~\ref{fig:IDE2p}) shifts the Hubble constant towards very larger values. The $H_0$ tension is then reduced within three standard deviations for all the combination of data sets considered in this work.

Both the interaction parameters $\xi_0$ and $\xi_a$ have only a lower limit for all the dataset combinations at 68\% CL and are consistent with zero, i.e. consistent  with a model without interaction, as we notice from Tab.~\ref{tab:resultsIDE2p}. A strong evidence for a phantom equation of state $w_x<-1$ is present at more than 2$\sigma$ for the CMB only case, and at many standard deviations for the CMB+R19 combination. However, this is not the case for CMB+BAO data. In this scenario {\rm IDE2p}, the $S_8$ value shifts down enough to solve the tension with the cosmic shear experiments for all the data combinations considered here.

Finally, for this IDE2p scenario, we have that the $\chi^2$ values are systematically higher than the $w$CDM model, as we can see in Table~\ref{tab:chi2}, showing that this is disfavoured by the fit of the data.

\begingroup                                          %\squeezetable     
\begin{center}                                       \begin{table*}                                       \begin{tabular}{cccccccccccccc}                         
\hline\hline                                                                                                                    
Parameters & Planck 2018 & Planck 2018+BAO & Planck 2018+R19 \\ \hline
$\Omega_c h^2$ & $    0.1203_{-       0.0027}^{+      0.0028}$ &  $    0.1206_{-       0.0023}^{+      0.0024}$ &   $    0.1208_{-       0.0027}^{+      0.0026}$  \\

$\Omega_b h^2$ & $    0.02235_{-      0.00030}^{+     0.00029}$ & $    0.02231_{-   0.00027}^{+     0.00027}$ &  $    0.02231_{-     0.00029}^{+    0.00029}$  \\

$100\theta_{MC}$ & $    1.04088_{-      0.00063}^{+      0.00062}$ & $    1.04086_{-    0.00060}^{+   0.00059}$ &  $    1.04083_{-      0.00061}^{+       0.00061}$  \\

$\tau$ & $    0.055_{-       0.015}^{+      0.016}$ & $    0.055_{-       0.015}^{+     0.016}$ &  $    0.055_{-     0.016}^{+      0.016}$  \\

$n_s$ & $    0.9639_{-      0.0088}^{+     0.0083}$ &  $    0.9629_{-        0.0079}^{+      0.0080}$ & $    0.9627_{-      0.0082}^{+        0.0086}$  \\

${\rm{ln}}(10^{10} A_s)$ & $    3.046_{-      0.031}^{+     0.032}$ & $    3.048_{-       0.031}^{+       0.032}$ &  $    3.047_{-        0.031}^{+      0.034}$   \\

$w_x$ & $   -1.67_{-      0.37}^{+     0.48}$ &  $ >-1.173$ &  $   -1.25_{-      0.10}^{+      0.10}$  \\

$\xi_0$ & $>-0.65$ &  $>-0.41$ &  $>-0.49$  \\

$\xi_a$ & ${\rm unconstrained}$ &  $>-0.72$ & $>-0.85$  \\

$\Omega_{m0}$ & $    0.186_{-       0.055}^{+       0.084}$ &  $    0.298_{-     0.022}^{+      0.021}$ &  $    0.260_{-    0.019}^{+       0.021}$  \\

$\sigma_8$ & $    0.93_{-        0.13}^{+       0.12}$ &  $    0.797_{-     0.055}^{+      0.048}$ &  $    0.834_{-       0.060}^{+       0.055}$  \\

$H_0 {\rm [km/s/Mpc]}$ & $>73$ &  $   69.4_{-       2.3}^{+       2.6}$ &  $   74.4_{-      2.7}^{+      2.8}$  \\

$S_8$ & $    0.725_{-       0.076}^{+       0.081}$ &  $    0.795_{-       0.050}^{+    0.045}$ &  $    0.776_{-        0.057}^{+      0.052}$  \\
\hline\hline                                                    \end{tabular}                                               \caption{95\% CL constraints on the interacting scenario {\rm IDE2p} using CMB from Planck 2018, BAO and local measurements of $H_0$ from R19. }
\label{tab:resultsIDE2p}                             \end{table*}                                         \end{center}                                         
\endgroup 
\begingroup  

\subsubsection{{\rm IDE2p} $+$ $M_{\nu}$ $+$ $N_{\rm eff}$}

The results for the {\rm IDE2} model within the phantom regime with the addition of the neutrino parameters, i.e. $M_{\nu}$ and $N_{\rm eff}$, are shown in Tab.~\ref{tab:resultsIDE2pnu} and Fig.~\ref{fig:IDE2pnu}.

As in the {\rm IDE1} model, the results from the previous section are not modified significantly with the introduction of $M_\nu$ and $N_{\rm eff}$ as extra parameters. Indeed, in this scenario the bound on $\Omega_ch^2$ is really robust, shifted only one standard deviation towards lower values with respect to the case in which the neutrino parameters are fixed, but still in agreement with what obtained in a $\Lambda$CDM model, see e.g. Tabs.~\ref{tab:resultsIDE2p} and \ref{tab:resultsIDE2pnu}. Also here the Hubble constant is almost unconstrained when the CMB data only is considered, due to the negative correlation with $w_x$, see Fig.~\ref{fig:IDE2pnu}. For the very same reason, the $H_0$ tension is reduced within 2.5$\sigma$ even after including BAO data in the analysis.

The neutrino sector parameters $M_\nu$ and $N_{\rm eff}$ do not show any strong correlation with the other cosmological parameters, with the exception of $w_x$, that is anti-correlated with the total neutrino mass $M_\nu$. As already pointed out, this anti-correlation between $w_x$ and $M_{\nu}$ is independent of the coupling in the dark sector. 
The preference for $w_x<-1$ is the reason of the softening of the $M_\nu$ upper limit. The most stringent bound we find on the sum of the neutrino masses is when adding BAO data to the CMB, i.e. $M_\nu<0.181$ eV at 95\% CL. The mean values of the effective number of relativistic degrees of freedom $N_{\rm eff}$ are lower than in a model without interaction $\xi(a)$, even if always highly consistent with its expected value $N_{\rm eff}=3.045$.

In Table~\ref{tab:chi2} we can see that the $\chi^2$ value for Planck 2018 data for this scenario (i.e., IDE2p $+$ $M_{\nu}$ $+$ $N_{\rm eff}$) is larger than the corresponding $\chi^2$ value obtained for the $w$CDM $+$ $M_{\nu}$ $+$ $N_{\rm eff}$ model, but concerning the other two datasets, the $\chi^2$ values for IDE2p $+$ $M_{\nu}$ $+$ $N_{\rm eff}$ are lower than the $w$CDM $+$ $M_{\nu}$ $+$ $N_{\rm eff}$ model. However, these lower values are consistent with the introduction of two more degrees, so do not correspond to an actual improvement of the fit. Therefore, these cases are almost equivalent.

\begingroup                                         % \squeezetable             
\begin{center}                                       \begin{table*}                                       \begin{tabular}{cccccccccccc}                        \hline\hline                                                                                                                    
Parameters & Planck 2018 & Planck 2018+BAO & Planck 2018+R19  \\ \hline
$\Omega_c h^2$ & $    0.1176_{-     0.0058}^{+   
0.0059}$ &  $    0.1179_{-      0.0057}^{+      0.0060}$ &  $    0.1176_{-     0.0057}^{+     0.0060}$ \\

$\Omega_b h^2$ & $    0.02215_{-       0.00045}^{+      0.00044}$ & $    0.02217_{-        0.00040}^{+        0.00040}$ &  $    0.02210_{-       0.00042}^{+       0.00043}$  \\

$100\theta_{MC}$ & $    1.04117_{-      0.00087}^{+      0.00089}$ & $    1.04118_{-        0.00086}^{+       0.00087}$ &  $    1.04119_{-       0.00088}^{+   
0.00087}$  \\

$\tau$ & $    0.054_{-   
0.015}^{+      0.016}$ & $    0.055_{-       0.015}^{+      0.016}$ &  $    0.053_{-        0.015}^{+       0.015}$ \\

$n_s$ & $    0.956_{-    0.017}^{+      0.016}$ & $    0.956_{-       0.015}^{+     0.015}$ &  $    0.954_{-     0.016}^{+      0.016}$  \\

${\rm{ln}}(10^{10} A_s)$ & $    3.036_{-       0.036}^{+      0.038}$ & $    3.039_{-    0.035}^{+      0.036}$ & $    3.036_{-     0.035}^{+      0.036}$  \\

$w_x$ & $   -1.76_{-        0.45}^{+     0.60}$ & $   >-1.22$ & $   -1.33_{-        0.20}^{+      0.18}$ \\

$\xi_0$ & $>-0.70$ & $>-0.43$ &  $>-0.52$ \\

$\xi_a$ & ${\rm unconstrained}$ & $ >-0.77$ & $ {\rm unconstrained}$ \\

$\Omega_{m0}$ & $    0.185_{-       0.057}^{+      0.089}$ & $    0.299_{-        0.024}^{+       0.021}$ & $    0.256_{-      0.022}^{+      0.023}$ \\

$\sigma_8$ & $    0.92_{-      0.14}^{+     0.13}$ & $    0.791_{-        0.058}^{+  0.050}$ &  $    0.825_{-      0.065}^{+       0.062}$ \\

$H_0 {\rm [km/s/Mpc]}$ & $>71$ & $   68.7_{-       3.1}^{+      3.3}$ &  $   74.2_{-      2.7}^{+       2.7}$ \\

$M_\nu {\rm [eV]}$ & $<0.365$ & $ <0.181$ &  $ <0.339$  \\

$N_{\rm eff}$ & $    2.84_{-        0.36}^{+       0.37}$ & $    2.86_{-     0.35}^{+      0.36}$ & $    2.82_{-      0.35}^{+      0.38}$   \\

$\Omega_\nu h^2$ & $<0.0038$ &  $<0.0019$ & $<0.0035$    \\

$S_8$ & $    0.714_{-      0.081}^{+      0.083}$ &  $    0.789_{-        0.053}^{+       0.048}$ &  $    0.762_{-       0.060}^{+      0.056}$  \\

\hline\hline                                         \end{tabular}                                        \caption{95\% CL constraints on the interacting scenario {\rm IDE2p} $+$ $M_{\nu}$ $+$ $N_{\rm eff}$ using CMB from Planck 2018, BAO and local measurements of $H_0$ from R19. }
\label{tab:resultsIDE2pnu}                           \end{table*}                                         \end{center}                                         
\endgroup 
\begin{figure*}
\includegraphics[width=0.66\textwidth]{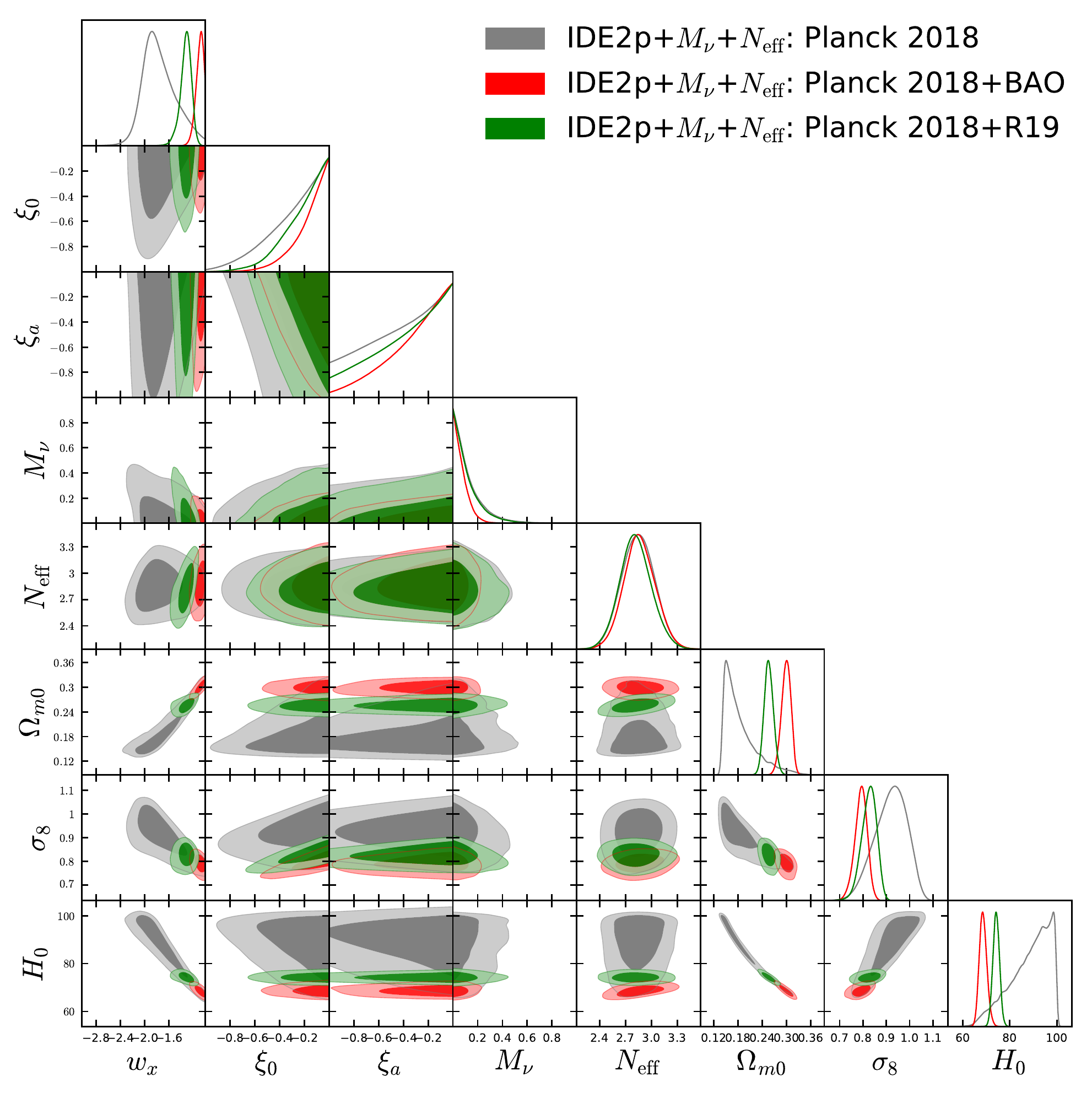}
\caption{One-dimensional marginalized posterior distributions and $68\%$ and $95\%$ CL two-dimensional contours for the interacting scenario {\rm IDE2p} $+$ $M_{\nu}$ $+$ $N_{\rm eff}$ for the  cosmological dataset combinations  considered  in this study.}
\label{fig:IDE2pnu}
\end{figure*}
\begingroup  

\subsubsection{{\rm IDE2q}}

The results for the {\rm IDE2} model in the quintessence regime, Eq.~(\ref{2q}), are presented in Tab.~\ref{tab:resultsIDE2q} and Fig.~\ref{fig:IDE2q}.

In the {\rm IDE2q} scenario, the well known anti-correlation present between $w_x$ and $H_0$, shifts the Hubble constant towards lower values, see Fig.~\ref{fig:IDE2q}, exacerbating the $H_0$ tension at more than 4$\sigma$ with respect to previous models.

Both the interaction parameters $\xi_0$ and $\xi_a$ are constrained by an upper limit for all the dataset combinations and are uncorrelated with the other cosmological parameters, as can be noticed from Fig.~\ref{fig:IDE2q}.
Only an upper limit is present also for the equation of state in the quintessence regime $w_x>-1$, and the $S_8$ tension with the cosmic shear experiments is restored.

Finally, even this IDE2q scenario is disfavoured by the fit of the data as showed in Table~\ref{tab:chi2}.

\begingroup
%\squeezetable        
\begin{center}                                       \begin{table*}                                       \begin{tabular}{ccccccccccccccc}     
\hline\hline                                                                                                                    
Parameters & Planck 2018 & Planck 2018+BAO & Planck 2018+R19 \\ \hline
$\Omega_c h^2$ & $    0.1200_{-        0.0027}^{+      0.0026}$ & $    0.1188_{-    0.0022}^{+       0.0022}$ & $    0.1175_{-   0.0025}^{+      0.0025}$   \\

$\Omega_b h^2$ & $    0.02238_{-        0.00029}^{+     0.00030}$ &  $    0.02246_{-      0.00028}^{+     0.00028}$   & $    0.02258_{-        0.00028}^{+   0.00029}$  \\

$100\theta_{MC}$ & $    1.04094_{-       0.00062}^{+     0.00060}$ & $    1.04108_{-   0.00060}^{+     0.00062}$  & $    1.04124_{-       0.00058}^{+        0.00060}$  \\

$\tau$ & $    0.053_{-       0.015}^{+       0.016}$ &  $    0.054_{-     0.015}^{+    0.015}$  & $    0.057_{-      0.015}^{+       0.016}$  \\

$n_s$ & $    0.9659_{-      0.0085}^{+       0.0089}$ &  $    0.9689_{-      0.0079}^{+     0.0077}$  & $    0.9720_{-       0.0082}^{+       0.0082}$  \\

${\rm{ln}}(10^{10} A_s)$ & $    3.042_{-        0.032}^{+        0.033}$ & $    3.042_{-     0.032}^{+     0.030}$  & $    3.043_{-       0.033}^{+    0.033}$   \\

$w_x$ & $<-0.79$ &  $<-0.925$  & $<-0.975$  \\

$\xi_0$ & $<0.159$ &  $<0.195$  & $<0.224$  \\

$\xi_a$ & $<0.36$ &  $<0.37$   & $<0.44$  \\

$\Omega_{m0}$ & $    0.337_{-       0.039}^{+       0.049}$ &  $    0.316_{-      0.016}^{+     0.017}$  & $    0.302_{-       0.015}^{+       0.015}$ \\

$\sigma_8$ & $    0.807_{-      0.054}^{+     0.042}$ & $    0.820_{-       0.033}^{+       0.036}$  & $    0.827_{-    0.031}^{+       0.0326}$   \\

$H_0 {\rm [km/s/Mpc]}$ & $   65.2_{-       4.3}^{+      3.3}$ &  $   67.1_{-    1.6}^{+      1.5}$ & $   68.3_{-      1.2}^{+     1.2}$  \\

$S_8$ & $    0.855_{-       0.037}^{+     0.038}$ &  $    0.841_{-       0.034}^{+      0.036}$  & $    0.830_{-      0.039}^{+    0.040}$  \\
\hline\hline                                         \end{tabular}                                        \caption{95\% CL constraints on the interacting scenario {\rm IDE2q} using CMB from Planck 2018, BAO and local measurements of $H_0$ from R19.  }
\label{tab:resultsIDE2q}                             \end{table*}                                         \end{center}                                         
\endgroup 
\begin{figure*}
\includegraphics[width=0.55\textwidth]{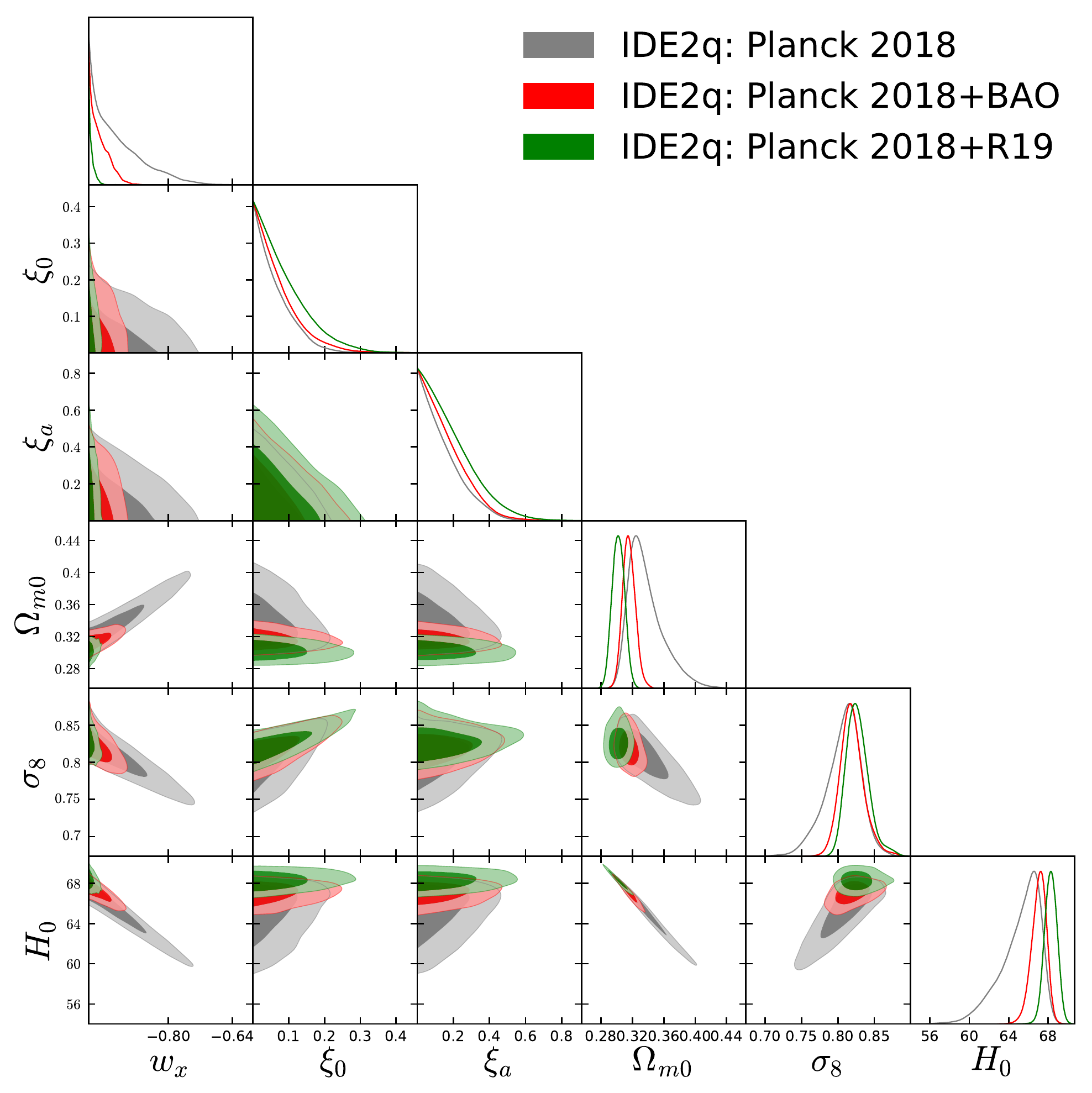}
\caption{One-dimensional marginalized posterior distributions and $68\%$ and $95\%$ CL two-dimensional contours for the interacting scenario {\rm IDE2q} for the  cosmological dataset combinations  considered  in this study.}
\label{fig:IDE2q}
\end{figure*}

\subsubsection{{\rm IDE2q} $+$ $M_{\nu}$ $+$ $N_{\rm eff}$}

The results for the {\rm IDE2} model in the quintessence regime with  $M_{\nu}$ plus $N_{\rm eff}$ as additional parameters are shown in Tab.~\ref{tab:resultsIDE2qnu} and Fig.~\ref{fig:IDE2qnu}.
However, for this model {\rm IDE2q}, the neutrino parameters $M_\nu$ and $N_{\rm eff}$ are  correlated with other cosmological parameters. In particular, we notice an important correlation with the Hubble constant $H_0$. This degeneracy is responsible, when the R19 prior is included in the data, i.e. for the combination CMB+R19, of the shift of $N_{\rm eff}$ towards higher values. The value $N_{\rm eff} = 3.43 ^{+0.15}_{-0.16}$ at 68\% CL deviates from the canonical expectation more than 2 standard deviations. In this {\rm IDE2q} scenario we  obtain our strongest limit on the $M_\nu$, $M_\nu<0.116$ eV at 95\% CL, as expected in quintessential non-interacting scenarios~\cite{Vagnozzi:2018jhn} where an anti-correlation between $w_x$ and $M_{\nu}$ exists similar to this coupled case. 

In Table~\ref{tab:chi2} we can see that the $\chi^2$ values for IDE2q $+$ $M_{\nu}$ $+$ $N_{\rm eff}$ are larger than the $\chi^2$ values obtained in the $w$CDM $+$ $M_{\nu}$ $+$ $N_{\rm eff}$ model for Planck 2018 alone and Planck 2018 + R19, but lower for Planck 2018 + BAO. However, this improvement quantified through $\Delta \chi^2 \sim 2$ is consistent with the the fact that in the interacting scenario we have two extra degrees of freedom, so it does not correspond to an actual improvement of the fit. 

\begingroup                                          %\squeezetable        
\begin{center}                                       \begin{table*}                                       \begin{tabular}{ccccccccccccccccccccc}               \hline\hline                                                                                                                    
Parameters & Planck 2018 & Planck 2018+BAO & Planck 2018+R19\\ \hline
$\Omega_c h^2$ & $    0.1187_{-      0.0060}^{+       0.0060}$ & $    0.1190_{-       0.0061}^{+      0.0060}$  & $    0.1237_{-      0.0053}^{+    0.0057}$ \\

$\Omega_b h^2$ & $    0.02225_{-    0.00046}^{+      0.00046}$ & $    0.02247_{-       0.00039}^{+     0.00038}$ & $    0.02284_{-        0.00034}^{+        0.00034}$  \\

$100\theta_{MC}$ & $    1.04105_{-       0.00088}^{+     0.00091}$ & $    1.04107_{-       0.00084}^{+      0.00089}$ & $    1.04057_{-        0.00079}^{+  0.00076}$  \\

$\tau$ & $    0.053_{-     0.015}^{+      0.016}$ & $    0.055_{-        0.015}^{+       0.016}$   & $    0.058_{-       0.016}^{+      0.016}$ \\

$n_s$ & $    0.961_{-       0.018}^{+    0.018}$ & $    0.969_{-        0.015}^{+      0.015}$  & $    0.985_{-       0.013}^{+     0.013}$ \\

${\rm{ln}}(10^{10} A_s)$ & $    3.037_{-     0.036}^{+       0.038}$ & $    3.042_{-      0.036}^{+      0.037}$  & $    3.060_{-     0.034}^{+     0.036}$  \\

$w_x$ & $<-0.77$ & $<-0.915$  & $<-0.965$   \\

$\xi_0$ & $  <0.17$ & $<0.17$ & $ <0.23$  \\

$\xi_a$ & $<0.39$ & $<0.39$  & $<0.48$  \\

$\Omega_{m0}$ & $    0.353_{-      0.056}^{+        0.068}$ & $    0.315_{-    0.016}^{+      0.018}$  & $    0.294_{-      0.016}^{+     0.018}$  \\

$\sigma_8$ & $    0.788_{-     0.075}^{+     0.066}$ & $    0.822_{-       0.036}^{+      0.037}$  & $    0.852_{-     0.038}^{+       0.040}$  \\

$H_0 {\rm [km/s/Mpc]}$ & $   63.7_{-     5.8}^{+     5.3}$ & $   67.2_{-   2.5}^{+     2.4}$  & $   70.7_{-   2.1}^{+      2.2}$  \\

$M_\nu {\rm [eV]}$ & $<0.41$ & $<0.137$  & $ <0.116$ \\

$N_{\rm eff}$ & $    2.94_{-     0.38}^{+      0.39}$ & $    3.06_{-     0.38}^{+        0.36}$  & $    3.43_{-      0.29}^{+    0.15+    0.32}$  \\

$\Omega_\nu h^2$ & $ <0.0043$ & $<0.00150$  & $ <0.00129$  \\

$S_8$ & $    0.853_{-      0.038}^{+     0.040}$ & $    0.842_{-       0.033}^{+      0.035}$ & $ 0.844_{-      0.037}^{+      0.044}$  \\

\hline\hline                                         \end{tabular}                                        \caption{95\% CL constraints on the interacting scenario {\rm IDE2q} $+M_{\nu}+N_{\rm eff}$ using CMB from Planck 2018, BAO and local measurements of $H_0$ from R19. }
\label{tab:resultsIDE2qnu} 
\end{table*}                                         \end{center}                                         
\endgroup                                                       
\begin{figure*}
\includegraphics[width=0.68\textwidth]{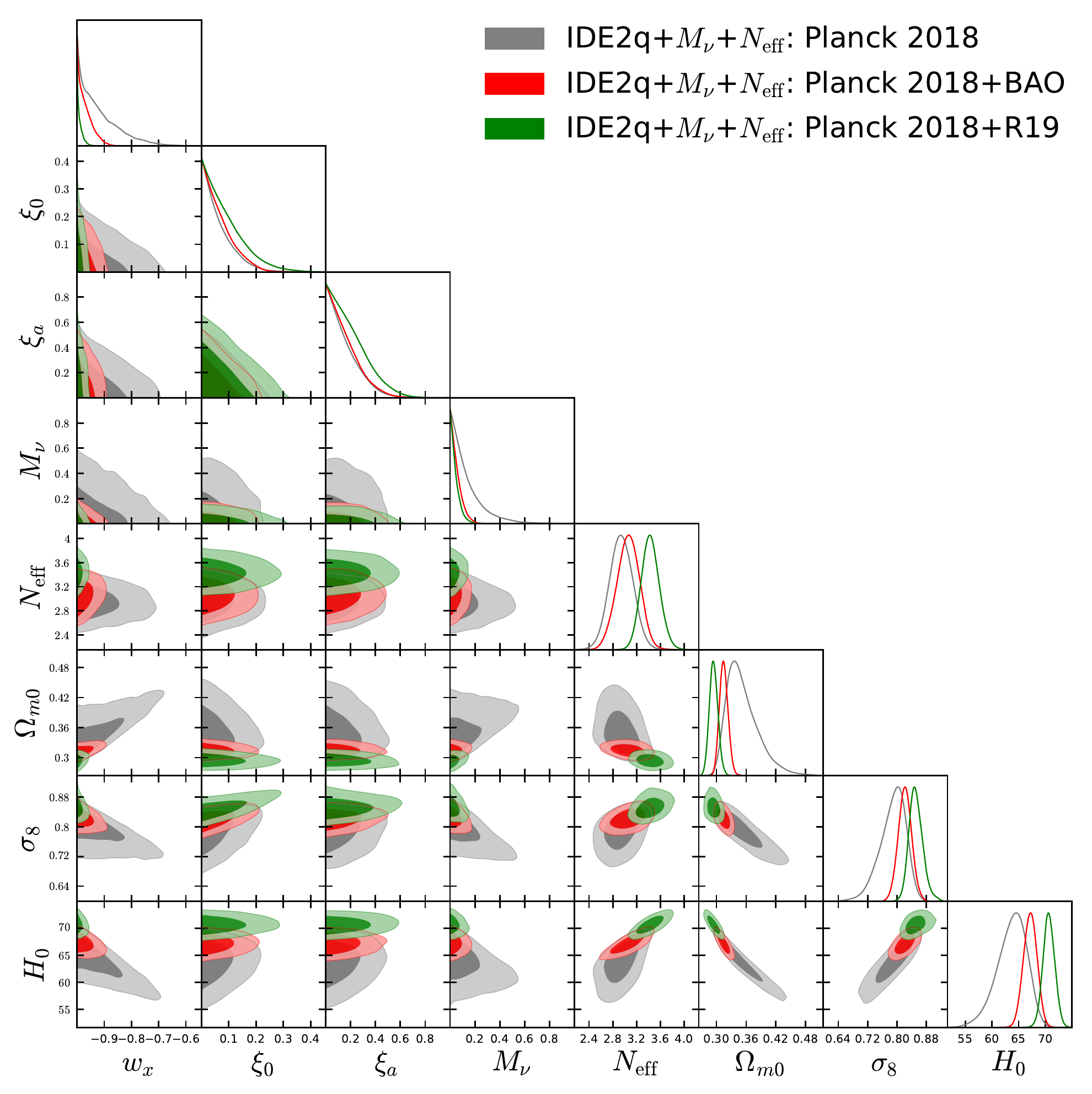}
\caption{One-dimensional marginalized posterior distributions and $68\%$ and $95\%$ CL two-dimensional contours for the interacting scenario {\rm IDE2q} $+$ $M_{\nu}$ $+$ $N_{\rm eff}$ for the  cosmological dataset combinations  considered  in this study.}
\label{fig:IDE2qnu}
\end{figure*}

%%%%%%%%%%%%%%%%%%%%%%%%%%%%%%%%%%%%%%%%%%%%%
\begingroup                                         %\squeezetable    
\begin{center}
\begin{table*} 
\begin{tabular}{cccccccccccccc} 
\hline\hline 
Parameters & Planck 2018 & Planck 2018+BAO & Planck 2018+R19 \\ \hline
$w$CDM & $ 2767.124$ & $2777.664$  & $ 2771.262$  \\
{\rm IDE1p} & $    2767.166$ & $    2775.306$ & $    2766.776$   \\
{\rm IDE1q} & $    2775.446$ & $    2780.372$  & $    2774.392$  \\
{\rm IDE2p} & $    2769.308$ & $    2781.104$  & $    2773.456$  \\
{\rm IDE2q} & $    2773.834$ & $    2779.528$  & $    2790.99$ \\
\hline
$w$CDM $+$ $M_{\nu}$ $+$ $N_{\rm eff}$ & $    2768.422$ & $    2779.370$  & $    2771.166$  \\
{\rm IDE1p} $+$ $M_{\nu}$ $+$ $N_{\rm eff}$ & $    2766.376$ & $    2776.278$ & $    2766.948$  \\
{\rm IDE1q} $+$ $M_{\nu}$ $+$ $N_{\rm eff}$ & $    2773.570$ & $    2779.448$  & $    2774.210$  \\
{\rm IDE2p} $+$ $M_{\nu}$ $+$ $N_{\rm eff}$ & $2769.558$ & $   2777.012$  & $   2770.830$ \\
{\rm IDE2q} $+$ $M_{\nu}$ $+$ $N_{\rm eff}$ & $2775.076$ & $2777.986$  & $2782.564$  \\

\hline\hline
\end{tabular} 
\caption{Best fit $\chi^2$ for the cases analysed here and the comparison with $w$CDM and $w$CDM $+$ $M_{\nu}$ $+$ $N_{\rm eff}$ models.}
\label{tab:chi2}
\end{table*}                                         \end{center}                                        
\endgroup  
%%%%%%%%%%%%%%%%%%%%%%%%%%%%%%%%%%%%%%%%%%%%%

\section{Summary and Conclusions}
\label{sec-conclu}

In this manuscript we further investigate the presence of a exchange rate $Q$ between dark matter (DM) and dark energy (DE) allowing for a time-dependent coupling~\cite{Yang:2019uzo}. We add new ingredients in the models, such as (i) freely varying neutrino parameters and (ii) a DE with a constant, freely varying equation-of-state, rather than vacuum dark energy. 
We restrict ourselves to the natural form of the coupling parameter $\xi (a) = \xi_0 + (1-a)\xi_a$ and consider two interacting models, namely, IDE1 ($Q = 3 H [\xi_0 + \xi_a (1-a)] \rho_x$) and IDE2 ($Q = 3 H [\xi_0 + \xi_a (1-a)] \frac{\rho_c \rho_x}{\rho_c+\rho_x} $). In order to avoid instabilities in the perturbation evolution, we consider the  regions (A) $w_x< -1,\; \xi_0 < 0,\; \xi_a < 0$, and (B) $w_x> -1,\; \xi_0 > 0,\; \xi_a > 0$, and investigate the interacting scenarios with and without the presence of neutrinos. The scenario with phantom DE equation-of-state ($w_x< -1$) is labeled as {\rm IDEp} and the scenario where DE has a quintessence-like equation-of-state ($w_x> -1$) is labeled as {\rm IDEq}. 
Let us summarize the main observational results that we find for all these scenarios:

\begin{itemize}
    \item {\rm IDE1:} We have explored this interaction model for both regimes, namely, $w_x< -$ and $w_x> -1$ with and without the presence of neutrinos. We have therefore investigated four different scenarios: {\rm IDE1p}, {\rm IDE1p} $+$ $M_{\nu}$ $+$  $N_{\rm eff}$, {\rm IDE1q}, and {\rm IDE1q} $+$  $M_{\nu}$ $+$  $N_{\rm eff}$.
    
    We find that for both {\rm IDE1p} and {\rm IDE1p} $+ M_{\nu}+ N_{\rm eff}$, $\Omega_ch^2$ is larger than within the $\Lambda$CDM cosmology and $w_x$ prefers a phantom nature with high significance. The parameter $\xi_0$ determining the current value of the DM-DE interaction is consistent with a null value, while $\xi_a$ prefers a value different from zero (albeit only mildly). We also notice that within these two phantom frameworks, the tension on $H_0$ is alleviated satisfactorily for all the data combinations considered here (CMB, CMB+BAO and CMB+R19). Concerning the $S_8$ parameter, its tension is significantly reduced only for the case of CMB data alone. The inclusion of $M_{\nu}$ and $N_{\rm eff}$ to {\rm IDE1p} does not change the constraints on other parameters.  The most stringent bound on $M_{\nu}$ is obtained for the CMB+BAO case and is $M_{\nu }< 0.162$~eV~ at  95\% CL.  
    
    As regards the remaining two scenarios {\rm IDE1q} and {\rm IDE1q} $+$ $M_{\nu}$ $+$ $N_{\rm eff}$, similarly to the phantom case, the inclusion of the neutrinos does not affect the constraints on the remaining cosmological parameters. The tightest bound on $M_{\nu} $ appears for CMB+BAO case ($M_{\nu }< 0.189$~eV at 95\% CL) which is slightly larger than the one obtained within the $\Lambda$CDM framework for the same data combination. Contrarily to the previous two cases, the value of  $\Omega_c h^2$ is much smaller. The parameter  $\xi_0$ is found to be non-zero for all the cases. However, $\xi_a$ is consistent with zero for all the datasets exploited in this work. The $H_0$ tension is solved for the CMB case ($H_0 = 70.2^{+4.1}_{-3.1}$ km/s/Mpc) and due to the very large error bars on the $S_8$ parameter, the $S_8$ tension is mildly alleviated.

    \item {\rm IDE2:} Using the very same observational data than for {\rm IDE1}, we have investigated four scenarios, namely, {\rm IDE2p}, {\rm IDE2p} $+$ $M_{\nu}$ $+$ $N_{\rm eff}$, {\rm IDE2q}, and {\rm IDE2q} $+$  $M_{\nu}$ $+$  $N_{\rm eff}$.  
    
    The scenario {\rm IDE2p} is very interesting because both the $H_0$ and $S_8 $ tensions are alleviated for all the data combinations used in this analysis. The dark energy equation of state shows a strong preference for  a  phantom nature. When neutrinos are considered into this picture ({\rm IDE2p} $+ M_{\nu}+ N_{\rm eff}$) no significant changes are obtained, apart from  the large anti-correlation between $w_x$ and $M_{\nu}$. The DE equation of state still prefers $w_x< -1$ with high significance. Finally, for both {\rm IDE2p} and {\rm IDE2p} $+$ $M_{\nu}$ $+$ $N_{\rm eff}$ models we find that $\xi_0$ and $\xi_a$ are consistent with zero, leading to a negligible preference for an interacting scenario. 
    
    The scenario {\rm IDE2q} is quite different from the previous cases. Within this interaction scheme we find that none of the tensions ($H_0$,  $S_8$) are alleviated. We do not find any evidence for an interaction among the dark sectors, since both the parameters $\xi_0$ and $\xi_a$, quantifying the interaction, are consistent with zero. An interesting outcome of this scenario is that it provides the most stringent bound on $M_{\nu}$ found in this study ($M_{\nu }< 0.116$ eV at 95\% CL), which is obtained for the combination of CMB+R19.

\end{itemize}
Finally, to conclude, 
the bounds on the effective number of neutrino species $N_{\rm eff}$ as we see in almost all of the scenarios above are extremely robust and consistent with the standard value of $N_{\rm eff} = 3.046$ and are, therefore, completely unaffected by the dynamics of the dark sectors. However, the $H_0$ tension scenario which for some cases in this work is alleviated, needs further investigations in light of other cosmological datasets. The excess of lensing in the CMB damping tail (see for instance \cite{Motloch:2018pjy}) might be an appealing investigation in this context.

\acknowledgments 
The authors thank the referee for some valuable comments aiming to improve the quality of the manuscript. 
WY acknowledges the support from the National Natural Science Foundation of China under Grants No.  
11705079 and No.  11647153. EDV was supported from the European Research Council in the form of a Consolidator Grant 
with number 681431. OM is supported  by  the  Spanish  grants FPA2017-85985-P and SEV-2014-0398 of the MINECO, by PROMETEO/2019/083 and by the European Union Horizon  2020  research  and  innovation  program  (grant  agreements No.  690575 and 67489). SP was supported by the the Science and Engineering Research Board (SERB), Govt. of India through the Mathematical Research Impact-Centric Support Scheme (MATRICS), File No. MTR/2018/000940.

\end{document}